# Interlayer-engineered local epitaxial templating induced enhancement in polarization (2P$_r$>70μC/cm²) in Hf$_{0.5}$Zr$_{0.5}$O$_2$ thin films

Srinu Rowtu, Paritosh Meihar, Adityanarayan Pandey, Md. Hanif Ali, Sandip Lashkare, Udayan Ganguly, *Senior Member, IEEE*

*Abstract*— In this work, we report a high remnant polarization, 2P$_r$ >70 μC/cm² in thermally processed atomic layer deposited Hf$_{0.5}$Zr$_{0.5}$O$_2$ (HZO) film on Silicon with NH$_3$ plasma exposed thin TiN interlayer and Tungsten (W) as a top electrode. The effect of interlayer on the ferroelectric properties of HZO is compared with standard Metal-Ferroelectric-Metal and Metal-Ferroelectric-Semiconductor structures. X-Ray Diffraction shows that the Orthorhombic (o) phase increases as TiN is thinned. However, the strain in the o-phase is highest at 2 nm TiN and then relaxes significantly for the no-TiN case. HRTEM images reveal that the ultra-thin TiN acts as a seed layer for the local epitaxy in HZO potentially increasing the strain to produce a 2X improvement in the remnant polarization. Finally, the HZO devices are shown to be wake-up-free, and exhibit endurance >10⁶ cycles. This study opens a pathway to achieve epitaxial ferroelectric HZO films on Si with improved memory performance.

*Index Terms*— Endurance, Ferroelectric memory, HZO, interlayer, local epitaxy, wake-up free

## I. Introduction

FERROELECTRIC HfO$_2$-based memory devices are promising candidates for non-volatile memory and neuromorphic applications because of their low power, excellent scalability, high switching speed and CMOS compatibility [1]–[3]. HfO$_2$ exhibits polymorphism at different temperatures. The ferroelectric properties associated with these materials are attributed to a meta-stable non-centrosymmetric polar Pca2$_1$ orthorhombic phase (o-phase) [4], which depicted a theoretical polarization of 50 μC/cm² and an experimentally observed coercive filed (E$_c$) of ~1-1.1 MV/cm [1]-[4]. The stabilization of o-phase in HfO$_2$ at lower temperatures requires certain conditions such as doping or cap-layer confinement over the other phases P2$_1$/c monoclinic and P4$_2$/nmc tetragonal. Various doping materials Si, Zr and others [5], cap-layer W, TiN and others, HfO$_2$ thickness and temperature variations have been extensively studied to obtain stable and reliable ferroelectricity in HfO$_2$ [6]–[10]. The HfO$_2$ and ZrO$_2$ solid solutions mix well and the mixture is stable for wide range x in Hf$_{1-x}$Zr$_x$O$_2$ due to low energy requirement (a few hundreds of meV) to replace Hf/Zr atom in ZrO$_2$/HfO$_2$ [11]. The physical and chemical properties of Hf and Zr are very similar. In ALD deposited Hf$_{1-x}$Zr$_x$O$_2$, the maximum remnant polarization found at x = 0.5 (Hf$_{0.5}$Zr$_{0.5}$O$_2$), whereas for other dopants the doping percentage is < 10% [8]. This allows the alternate deposition of Hf and Zr precursors (1:1 ratio) in Atomic Layer Deposition (ALD) system which leads to homogenous and reproducible films for mass production. The H$_{0.5}$Z$_{0.5}$O$_2$ (HZO) is extensively studied as the large remnant polarization (P$_r$) is achieved at lower thermal budget (400-600 °C) [12]–[14]. The P$_r$ value is one of the major metrics for characterizing the quality of ferroelectric film, specifically the o-phase. The high P$_r$ indicates larger fraction of o-phase and better-quality ferroelectric films, with reduced degradation effects such as imprint [15], drift in endurance cycling [16], for applications where multiple conductance states are needed. Many processes are reported in literature to achieve the high polarization such as high-pressure post metallization annealing [17], thermal quenching [18], interlayer (IL) engineering [19]–[21] and plasma treatment [22],[23] during fabrication.

This work is supported in part by DST Nano Mission, Ministry of Electronics and Information Technology (MeitY), and Department of Electronics, through the Nano-electronics Network for Research and Applications (NNETRA) project of Govt. of India. It was performed at IIT Bombay Nanofabrication Facility.

All the authors are with the Department of Electrical Engineering, Indian Institute of Technology Bombay, Mumbai, 400076, India (e-mail: srinurowtu@gmail.com, anbp.phy@gmail.com, udayan@ee.iitb.ac.in). This article available at https://doi.org/10.1109/TED.2023.3277804
Digital Object Identifier 10.1109/TED.2023.3277804

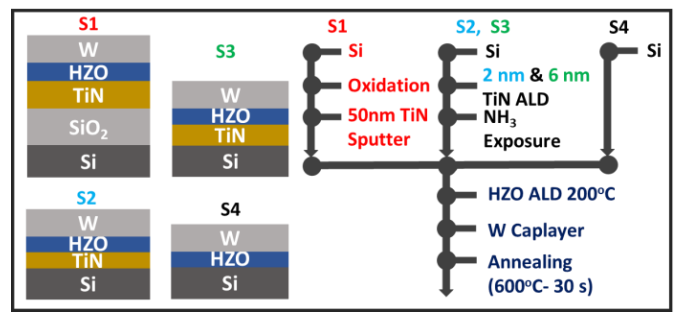

Fig. 1. Schematic diagram of 4 device stacks and process flow (a) S1: TiN (50 nm) /HZO/W where W and TiN are TE and BE, respectively,

(b) S2: Si/TiN(2 nm)/HZO/W, and (c) S3: Si/TiN(6 nm)/HZO/W (d) S4: Si/HZO/W. In S2, S3, high doped Si substrate is used as BE due to low TiN thickness. (d) S4: Si/HZO/W stack where Si substrate is BE.

In this study, we demonstrate the enhanced polarization in the HZO films by introducing a thin TiN interlayer (IL) and NH$_3$ plasma treatment. The Si/TiN (2 nm)/HZO/W stack with NH$_3$ exposure shows wakeup-free 2Pr >70 µC/cm$^2$ and endurance up to ~6x10$^6$ cycles. We demonstrate that TiN IL plays a crucial role in the crystallization of the o-phase in HZO. The local epitaxy is observed in HRTEM images for a 2 nm TiN IL stack which leads to larger domains and subsequently enhancement of the ferroelectric properties of HZO.

## II. EXPERIMENT

In this study, 4 types of ferroelectric capacitors are compared in Fig. 1. A standard MFM stack on Si/SiO$_2$ substrate (Si/SiO$_2$/TiN/HZO/W) is fabricated namely S1. A 50 nm TiN film is deposited by using sputter. The MFS stacks, with TiN as interlayer (Si/TiN/HZO/W), are fabricated on heavily doped Si (100) substrate (0.001-0.005 ohm-cm) with 2 nm (S2), 6 nm (S3) thickness of TiN and without TiN (S4). For device stacks S2 and S3, TiN is deposited by ALD at 250 °C with Ti[(CH$_3$)$_2$N]$_4$ and NH$_3$ as precursors, after which samples are exposed to NH$_3$ plasma. The Ferroelectric layer HfO$_2$: ZrO$_2$ (~10 nm) layer deposited by ALD with alternate cycles of HfO$_2$ and ZrO$_2$ at 200 °C. Hf[(CH$_3$)$_2$N]$_4$, Zr[(CH$_3$)$_2$N]$_4$, and H$_2$O are the precursors used for Hf, Zr, and O correspondingly. The top electrode, W (~50 nm), is patterned and deposited by sputtering followed by a lift-off process. Finally, the stacks are annealed in N$_2$ ambient at 600 °C for 30 s by using the rapid thermal annealing (RTA) process.

The structural characteristics of HZO film are investigated using grazing incidence X-ray diffraction (GIXRD) by high-resolution Rigaku Smartlab system with the Cu-K$\alpha$ radiation using 5 mm x 5 mm slit at a fixed incidence angle ω = 0.5º for a 2θ range of 20°-70° and thickness of the film is confirmed by X-ray reflectivity (XRR) measurement. For the GIXRD measurement, a blanket W metal is deposited covering the entire HZO film and W is etched using 30% H$_2$O$_2$ after annealing. High-resolution transmission electron microscopy (HRTEM) imaging of the films is carried out using Thermo Scientific Themis 300 G3 apparatus operating at 300 kV. The HRTEM images and its fast fourier transform are analysed using imageJ and CrysTBox tools [24]. The surface and depth X-ray photoelectron spectroscopy (XPS) scan is performed using ULVAC-PHI PHI5000 VersaProbeII utilizing Al-K$\alpha$ radiation as an excitation source for the elemental study. The Ar-ion etching is carried out for the XPS depth profile measurement. The XPS spectra is deconvoluted and analyzed using MultiPaK ESCA software. The electrical measurements are carried out by using Agilent B1500 semiconductor device analyzer with a B1530A waveform Generator module. The device dimensions used for the electrical measurements are 50 µm x 50 µm.

## III. RESULTS AND DISCUSSION

Fig. 2(a) shows the typical butterfly-like dielectric constant (ε$_r$)-E curves obtained from capacitance-voltage characteristics measured at 100 kHz frequency. The butterfly hysteresis in ε$_r$ indicates the ferroelectric nature of the HZO films. The S2 and S3 samples have higher dielectric constants (~52 and ~43) compared with the other two stacks. This is due to the enhancement of ferroelectric properties in these films. The well-defined ferroelectric PUND hysteresis loops performed at 3.5 MV/cm applied field are compared in Fig. 2(b). The PUND measurement is performed to deduce ferroelectric polarization arising from domain switching by subtracting leakage current and linear dielectric response. The obtained 2P$_r$ values are S1: 39.2 µC/cm$^2$, S2: 62.4 µC/cm$^2$, S3: 46.2 µC/cm$^2$, S4: 30.4 µC/cm$^2$ (Fig. 2(b)).

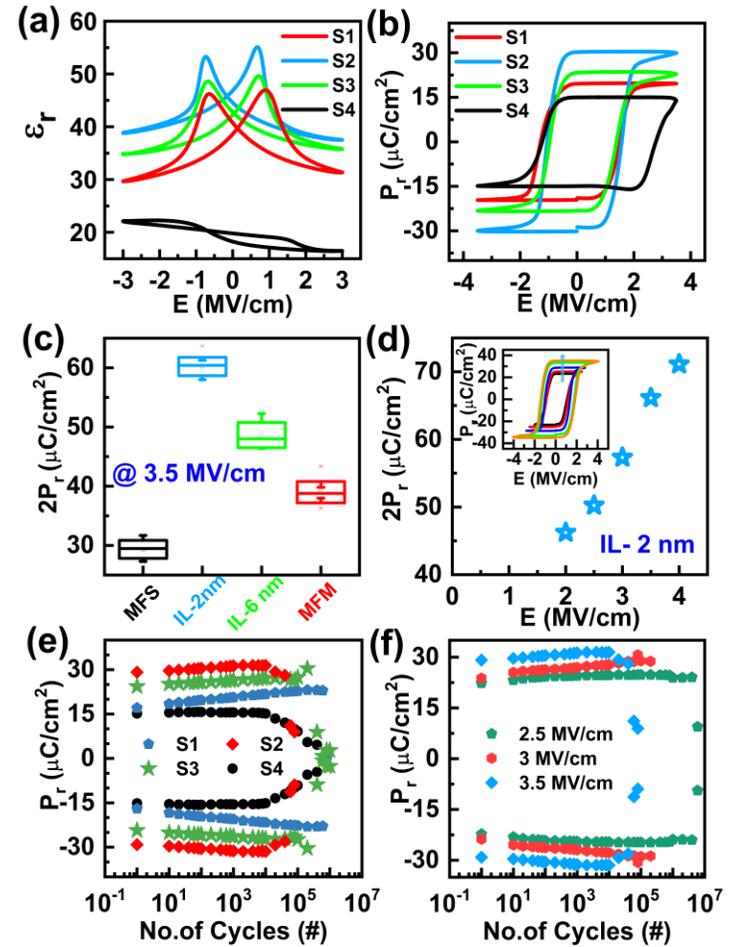

Fig. 2. (a) Dielectric constant Vs electric field curves obtained from C-V measurements at 100 kHz (b) PUND hysteresis loop of devices (c) statistical variation of 2P$_r$ among the 10 devices (d) Remnant polarization (2P$_r$) of TiN 2 nm NH$_3$ exposed device (S2) with increasing the electric field from 2 MV/cm V to 4 MV/cm. Inset shows the corresponding PUND hysteresis loops. (e) Endurance comparison among the splits at 3.5 MV/cm. (f) Endurance characteristics of S2- 2 nm TiN/HZO/W stack devices with 3.5, 3 and 2.5 MV/cm electric filed. Endurance cycles are >10$^6$ for 2.5 MV/cm.

Akin to the dielectric constant, the TiN IL and NH$_3$ exposed samples show higher polarization compared with the other two films. Fig. 2(c) shows a low (<5%) device-to-device (10 devices) variability of 2P$_r$ for all the samples. Further, the increment in polarization, for the S2 sample with increasing electric field is shown in Fig. 2(d). Here, the highest 2P$_r$ of 72 μC/cm$^2$ is demonstrated at 4 MV/cm. Endurance characteristics of all the stacks are shown in Fig. 2(e), stressing at 3.5 MV/cm with 40 μs triangular waveform. No wake-up effect is observed for S2, S3, and S4 devices, except in S1. The wake-up effect in TiN electrodes is well known and ammonia exposed TiN surface prevents the scavenging of oxygen from HZO and reduces the oxidation of TiN which will decrease the defect density at the TiN/HZO leads larger o-phase grains in HZO [22]-[23]. The endurance of the highest 2P$_r$ sample, S2, is limited to ~4x10$^4$ cycles, consequently, devices become leaky followed by hard breakdown. S1, S2, and, S3 samples exhibit breakdown limited endurance whereas S4 shows fatigue. Furthermore, the effect of the electric field on endurance is studied with different amplitudes for S2 from 2.5 MV/cm to 3.5 MV/cm as shown in Fig. 2(f). A trade-off is observed between field strength and endurance. The lower fields lead to less degradation, and enhance the endurance up to ~6x10$^6$ cycles, as opposed to higher fields. Further, GIXRD and HRTEM are performed to understand the origin of high polarization.

and t-phase (101)$_t$ of 2θ between these two phases in standard data and further distinguishing of these phases is done by using the deconvolution of the peak as shown in Fig. 3(c). A close inspection of o(111)/t(101)-peak depicts a shift towards lower 2θ for HZO films with interlayer in comparison to S1 and S4 in Fig. 3(a). This shift is due to an increase in the ratio of o- and t-phases which is also confirmed by peak deconvolution [20],[26]. The decrease of 2θ corresponds to an increase of the d$_{111}$ value from 2.92 (S4) to 2.94 Å (S2), as determined from the GIXRD (Fig. 3(b)). The sample S2 with IL 2 nm TiN has d$_{111}$ value of 2.94 Å which is close to the reference value of 2.95 Å calculated from the standard spectrum. The intensity ratio of o(200) and o(111) is carried out to determine the degree of the texture present in the film as shown in Fig. 3(b). The I$_{200}$/I$_{111}$ of S2 is high ~0.46 in comparison to other films and the polycrystalline simulated sample (~0.12) implying the presence of a large o(200) texture which is later confirmed in HRTEM investigation. The peaks around 30° are deconvoluted using the Gaussian shape function as shown in Fig 3(d). The relative phase fraction is calculated from the peak intensity and area under the curve methods by using the Gaussian and Lorentzian shape functions. The higher (0.91±0.04) o-phase obtained from deconvolution for 2 nm IL film in comparison to other films (S1 0.71±0.08, S3 ~0.77±0.02, and S4 ~0.69±0.05) indicate enhanced ferroelectric properties. The shift in 2θ towards the lower angle indicates the enhancement of the o-phase fraction in the S2 sample which results in enhanced ferroelectric properties compared with other device stacks [20].

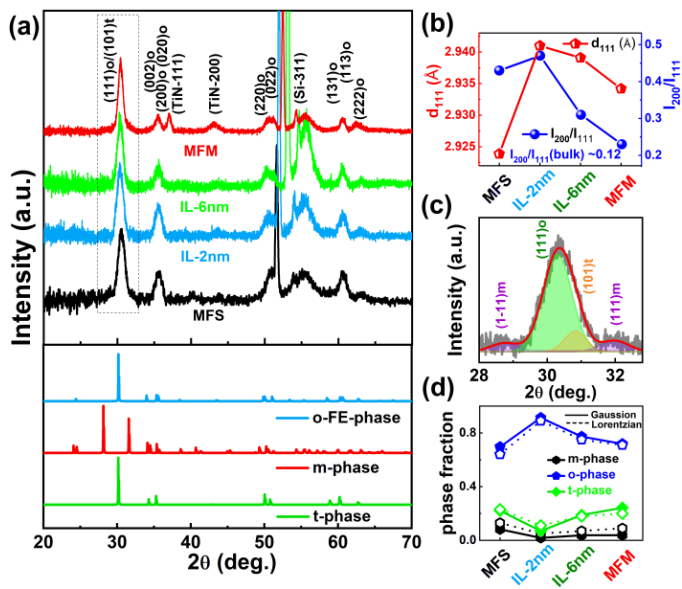

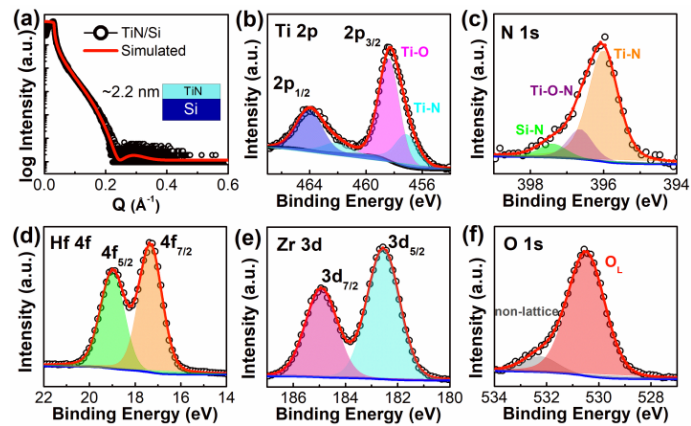

Fig. 3. GIXRD spectra of HZO crystallized films with incident angle 0.5° of S1, S2, S3, S4 device stacks for 2θ range of (a) 20°-70° shown with standard JCPDF database (m-phase: 01-075-6426, o-phase: 00-021-0904, t-phase: 00-008-0342). (b) The comparison of d$_{111}$ (Å) and intensity ratio of (200) and (111) peaks I$_{200}$/I$_{111}$ was calculated from the XRD spectra. (c) Deconvoluted GIXRD spectra of S2: (Si/TiN(2 nm)/HZO/W). (d) The relative phase fractions of the device stacks. o-phase intensity is 0.91±0.4 in S2. The suppression of m-phase and o/t-phase intensity is high for S2 compared with other stacks. GIXRD spectra measured after tope electrode W etching.

As shown in Fig. 3(a), all the films are polycrystalline in nature and crystallized to almost o-/t-phases with suppressed m-phase due to in-plane strain applied by electrodes during annealing. The peaks at ~28.5°, ~30.5° and ~31.6° are assigned to m (-111), o(111)/t(101) and m(111), respectively [25]. The peak at ~30.5° is assigned for both o-/t-phases due to closely overlapping diffraction reflection of o-phase (111)$_o$

Fig. 4. (a) XRR spectra of interlayer TiN (~2.2±0.2 nm) deposited on Si and fitted with single layer model to extract thickness, XPS spectra of TiN/Si stack (b) Ti 2p (c) N 1s depicting formation of Ti-N, Ti-O-N, and Si-N bonds, XPS spectra of S2 film (d) Hf 4f, (e) Zr 3d and (f) O 1s illustrating Hf-O/Zr-O bond formation.

The XPS measurement is performed to determine the bonding of the elements and the chemical composition of the film. Fig. 4(a)-(c) shows the XRR and XPS spectra along with fitting for interlayer TiN film deposited on Si. The estimated thickness of the TiN layer is ~2.2±0.2 nm. The deconvoluted XPS spectra of Ti 2p suggest the formation of Ti-O and TiN bonds and N 1s reveal Ti-N, Ti-O-N, and Si-N bond formation [27]. The high-resolution surface XPS analysis of S2 and S4 doesn't show a significant difference. Fig. 4(d)-(e) shows the XPS spectra of Hf 4f, Zr 3d, and O 1s for HZO film deposited on Si/TiN (2 nm), respectively. The Hf 4f and Zr 3d depict doublet peaks and the binding energy (BE) of Hf 4f$_{7/2}$ and Zr 3d$_{5/2}$ are located at 17.3 eV and 182.6 eV, respectively, revealing Hf-O/Zr-O bond formation [28]. The O 1s peak at ~530.6 eV is associated with the lattice oxygen and the higher BE ~532.3 eV corresponds to the non-lattice oxygen related to the overlap of surface OH groups [28]. The XPS depth profile analysis is carried out to confirm the stoichiometry as the earlier literature suggested that the ferroelectricity of the HZO film is also determined by the oxygen content in the film [29]. The XPS depth profile is performed for two stacks S4, and S2 as shown in Fig. 5(a) and (b), respectively. The presence of thin TiN layer near the interface between HZO and Si is confirmed for S2. It is clear that the Hf to Zr ratio is close to ~1:1 but films are not stoichiometric with oxygen content (Hf$_{0.5}$Zr$_{0.5}$O$_{1.9}$) revealing rich oxygen vacancy (V$_o$) for both stacks and the high V$_o$ is advantageous for the ferroelectricity in HZO films. Further, HRTEM imaging is carried out for the above two stacks to explicitly visualize the effect of interlayer on the crystallization of HZO film. Fig. 5(c) and (d) compare the cross-sectional HRTEM images of S4 and S2, respectively.

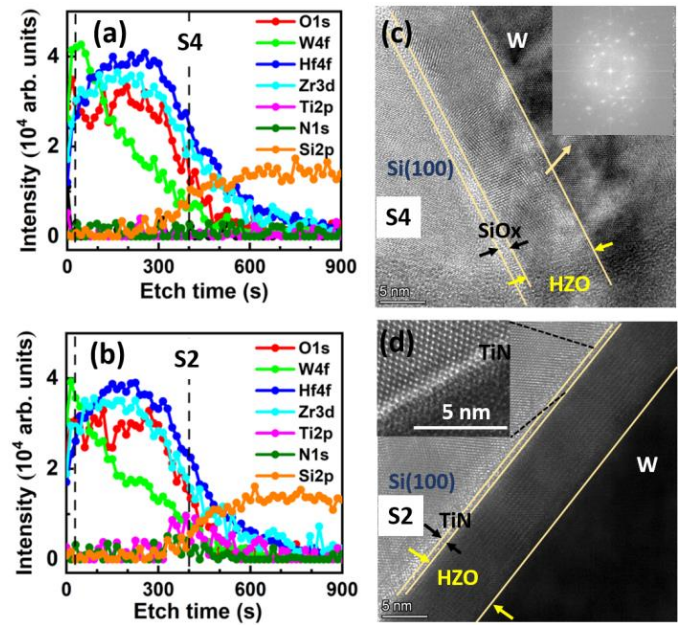

Fig. 5. XPS depth profile of (a) S4: Si/HZO/W, (b) S2: Si/TiN(2 nm)/HZO/W showing presence of TiN. Cross-sectional TEM image of (c) S4: Si/HZO/W, the right insets correspond to FFT images of the entire HZO layer. (d) S2: Si/TiN(2 nm)/HZO/W stacks, the left inset depicts the Si/TiN interface.

The well-formed W/HZO/Si structure and lattice fringes are clearly evident in both films. An amorphous SiOx layer (~2 nm) at Si/HZO interface with multiple crystallites in HZO can be seen for S4 in Fig. 5(c). The multiple crystallites are formed due to the formation of an amorphous interlayer between Si and HZO interface which is also confirmed from fast fourier transform (FFT). Whereas, in the case of S2, Si/HZO interlayer is ordered in a crystal pattern as shown in the left inset of Fig. 5(d). A close inspection of S2 (W/HZO/TiN(2 nm)/Si stack) revealed that the HZO layer shows the same crystallographic orientation as Si substrate. Fig. 6(a) shows the expanded view of TEM image of sample S2. Fig. 6(b) shows the real space schematic model of the epitaxial Si/TiN/HZO stack. The FFT of the entire HZO film illustrates zone axis similar to the selected area electron diffraction of Si substrate revealing the local epitaxial relation of HZO (100) film with Si (100) substrate along the [-110] zone axis (Fig. 6(c), 6(d)) of Si. The out-of-plane d-spacing ~5.05±0.01 Å matches with 'c' lattice constant (001) plane spacing) of HZO reported in the literature [30]. Also, the expanded view of a square region as shown in Fig. 6(d) is clearly matching with simulated (-110) surface of HZO.

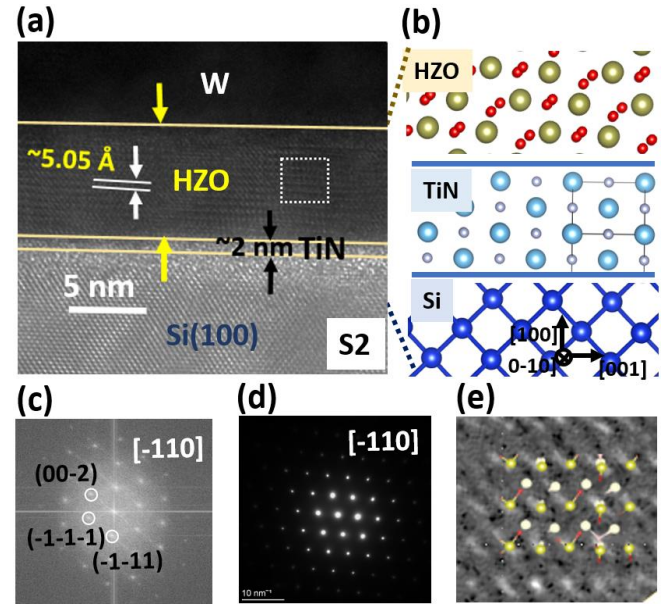

Fig. 6. (a) HRTEM image of S2 (Si/TiN (2 nm)/HZO/W) stack; (b) Real space schematic model of epitaxial Si/TiN/HZO stack (c) FFT images of entire HZO layer, (d) SAED pattern of Si substrate, the FFT of HZO film in S2 is similar to Si substrate revealing the local epitaxial relation along [-110] zone axis of Si, (e) expanded view of region framed by white square matched with simulated (-110) crystal surface of HZO.

The local epitaxy of HZO is possible due to the epitaxy of TiN on Si which is earlier reported in the literature and TiN

TABLE 1: BENCHMARKING WITH STATE OF THE ART WORK

| Ref | Device stack | Deposition Conditions | 2P$_r$ (µC/cm$^2$) | Wakeup cycles | Endurance |
|---|---|---|---|---|---|
| [7] | TiN/HZO/W | ALD 300° C | 58.7 | 10$^4$ | 10$^3$ |
| [7] | TiN/HZO/MO | ALD, 300 °C | 50.1 | - | - |
| [17] | TiN/HZO/TiN | PEALD, 300°C | 57.3 | Yes | - |
| [20] | W/HZO/W | PEALD, 250°C | 64 | 10$^3$ | 10$^4$ |
| [23] | TiN/HZO/TiN | ALD, 260°C | 60 | 10$^3$ | 10$^5$ |
| **This work** | **Si/TiN (2 nm)/HZO/W** | **ALD, 200°C** | **72** | **No** | **6x10$^6$ *** |

* Endurance measured at 2.5 MV/cm

acts as a seed layer for the formation of local epitaxial HZO crystallites [31]. The epitaxy of TiN on Si is reported in the literature [32]-[33] by PVD techniques. However, the TiN epitaxy is achieved here by using the solid phase epitaxy where the amorphous layer is sandwiched between the crystalline and caplayer. The TiN act as a buffer layer [34]-[35] between the Si and HZO film for epitaxy growth of HZO as shown in the real space schematic model Fig. 6(b). However, in the case of S4, due to the formation of an amorphous layer, the discontinuity between the Si/HZO interface and hence we observe multiple disordered crystallites.

Finally, as compared with the state-of-the-art, we demonstrate the highest 2P$_r$ of ~72 μC/cm$^2$ in HZO (Table 1). There are multiple factors contributing towards the highest 2P$_r$ in S2 (Si/TiN(2 nm)/HZO/W) device stack such as higher polar o-phase fraction, the texturing of HZO with c-axis, and the larger o-phase crystallites which arise from the local epitaxy of HZO. The presence of oxygen vacancies and mechanical stress by top electrode confinement with low thermal expansion coefficient metal W ~4.5 x 10$^{-6}$ /K, and the lattice mismatch between TiN and HZO are responsible for maximizing the o-phase fraction which leads to the enhancement in ferroelectricity [36]-[37]. The insertion of TiN seed layer and NH$_3$ plasma exposure minimize the defect density at the interfacial layer between TiN and HZO leading to local epitaxial HZO films on Si with improved ferroelectricity. This opens a new route to enhance the ferroelectric properties in HZO solid solution.

## IV. CONCLUSION

We showed a high 2P$_r$ ~72μC/cm$^2$ in Si/TiN-2 nm/HZO/W stack with the low thermal budget. The GIXRD and TEM revealed that the cause for enhanced polarization is the larger o-phase fraction in HZO films, caused by the strain from the electrodes and larger o-phase domains due to local epitaxy and less fraction of m-phase in HZO. The effect of thin TiN interlayer and NH$_3$ plasma exposure led to local epitaxy in HZO films due to the epitaxy in the TiN layer and it acts as a seed layer template between Si and HZO, confirmed through TEM analysis. The out-of-plane confinement with low TEC top electrode W metal and rich oxygen vacancies are responsible for a larger o-phase revealing high P$_r$. Further, the high polarization devices are wakeup-free and exhibit breakdown limited endurance up to ~6x10$^6$ cycles.

## REFERENCES

[1] H. W. Park, J. Lee, and C. S. Hwang, "Review of ferroelectric field-effect transistors for three-dimensional storage applications," *Nano Select*, vol. 2, no. 6, pp. 1187–1207, Jun. 2021, doi: 10.1002/nano.202000281.

[2] M. H. Park, Y. H. Lee, T. Mikolajick, U. Schroeder, and C. S. Hwang, "Review and perspective on ferroelectric HfO2-based thin films for memory applications," *MRS Communications*, vol. 8, no. 3, pp. 795–808, Sep. 2018. doi: 10.1557/mrc.2018.175.

[3] A. I. Khan, A. Keshavarzi, and S. Datta, "The future of ferroelectric field-effect transistor technology," *Nat Electron*, vol. 3, no. 10, pp. 588–597, Oct. 01, 2020. doi: 10.1038/s41928-020-00492-7.

[4] T. S. Böscke, J. Müller, D. Bräuhaus, U. Schröder, and U. Böttger, "Ferroelectricity in hafnium oxide thin films," *Appl. Phys. Lett.*, vol. 99, no. 10, Sep. 2011, doi: 10.1063/1.3634052.

[5] U. Schroeder, E. Yurchuk, J. Müller, D. Martin, T. Schenk, P. Polakowski, C. Adelmann, M. I. Popovici, S. V. Kalinin, and T. Mikolajick, "Impact of different dopants on the switching properties of ferroelectric hafniumoxide," in *Jpn. J.Appl. Phys.*, vol. 53, 2014, Art. no. 08LE02, doi: 10.7567/JJAP.53.08LE02.

[6] Z. Fan, J. Chen, and J. Wang, "Ferroelectric HfO2-based materials for next-generation ferroelectric memories," *J. Adv. Dielect.*, vol. 6, no. 2, 2016. doi: 10.1142/S2010135X16300036.

[7] Y. Lee, Y. Goh, J. Hwang, D. Das, and S. Jeon, "The Influence of Top and Bottom Metal Electrodes on Ferroelectricity of Hafnia," *IEEE Trans. Electron Devices*, vol. 68, no. 2, pp. 523–528, Feb. 2021, doi: 10.1109/TED.2020.3046173.

[8] H. A. Hsain, Y. Lee, M. Materano, T. Mittmann, A. Payne, T. Mikolajick, U. Schroeder, G. N. Parsons, and J. L. Jones, "Many routes to ferroelectric HfO$_2$: A review of current deposition methods," *J. Vac. Sci. Technol. A*40, 010803 (2022), doi: 10.1116/6.0001317.

[9] E. Yurchuk, J. Müller, S. Knebel, J. Sundqvist, A. P. Graham, T. Melde, U. Schroder, and T. Mikolakick, "Impact of layer thickness on the ferroelectric behaviour of silicon doped hafnium oxide thin films," in *Thin Solid Films*, 2013, vol. 533, pp. 88–92. doi: 10.1016/j.tsf.2012.11.125.

[10] M. H. Park, D. H. Lee, K. Yang, J. Y. Park, G. T. Yu, H. W. Park, M. Materano, T. Mittmann, P. D. Lomenzo, T. Mikolajick, U. Schroeder, and C. S. Hwang, "Review of defect chemistry in fluorite-structure ferroelectrics for future electronic devices," *J.Mater.Chem. C*, vol. 8, no. 31, pp. 10526–10550, Aug. 21, 2020. doi: 10.1039/d0tc01695k.

[11] C. Künneth, R. Materlik, M. Falkowski, and A. Kersch, "Impact of Four-Valent Doping on the Crystallographic Phase Formation for Ferroelectric HfO$_2$ from First-Principles: Implications for Ferroelectric Memory and Energy-Related Applications," ACS Appl. Nano Mater.1(1), 254-264 (2018). doi: 10.1021/acsanm.7b00124.

[12] M. H. Park, H. J. Kim, Y. J. Kim, T. Moon, and C. S. Hwang, "The effects of crystallographic orientation and strain of thin Hf$_{0.5}$Zr$_{0.5}$O$_2$ film on its ferroelectricity," *Appl. Phys. Lett.*, vol. 104, no. 7, Feb. 2014, doi: 10.1063/1.4866008.

[13] M. Hyuk Park, H. Joon Kim, Y. Jin Kim, W. Lee, T. Moon, and C. Seong Hwang, "Evolution of phases and ferroelectric properties of thin Hf$_{0.5}$Zr$_{0.5}$O$_2$ films according to the thickness and annealing temperature," *Appl. Phys. Lett.*, vol. 102, no. 24, Jun. 2013, doi: 10.1063/1.4811483.

[14] S. J. Kim, J. Mohan, S. R. Summerfelt, and J. Kim, "Ferroelectric Hf$_{0.5}$Zr$_{0.5}$O$_2$ Thin Films: A Review of Recent Advances," *JOM*, vol. 71, no. 1, pp. 246–255, Jan. 01, 2019. doi: 10.1007/s11837-018-3140-5.

[15] J. Bouaziz, P. Rojo Romeo, N. Baboux, and B. Vilquin, "Imprint issue during retention tests for HfO$_2$-based FRAM: An industrial challenge?," *Appl. Phys. Lett.*, vol. 118, no. 8, Feb. 2021, doi: 10.1063/5.0035687.

[16] S. Dünkel, M. Trentzsch, R. Richter, P. Moll, C. Fuchs, O. Gehring, M. Majer, S. Wittek, B. Müller, T. Melde, H. Mulaosmanovic, S. Slesazeck, S. Müller, J. Ocker, M. Noack, D.-A. Löhr, P. Polakowski, J. Müller, T. Mikolajick, J. Höntschel, B. Rice, J. Pellerin, S. Beyer, "A FeFET based super-low-power ultra-fast embedded NVM technology for 22nm FDSOI and beyond," in *IEDM. Tech. Dig.,* Dec. 2017, pp. 19.7.1-19.7.4, doi: 10.1109/IEDM.2017.8268425.

[17] D. Das, V. Gaddam, and S. Jeon, "Demonstration of High Ferroelectricity (Pr ~ 29 μc/cm2) in Zr Rich Hf$_x$Zr$_{1-x}$O$_2$ Films," *IEEE Electron Device Lett.*, vol. 41, no. 1, pp. 34–37, Jan. 2020, doi: 10.1109/LED.2019.2955198.

[18] B. Ku, S. Choi, Y. Song and C. Choi, "Fast Thermal Quenching on the Ferroelectric Al:HfO$_2$ Thin Film with Record Polarization Density and Flash Memory Application," in *Proc. IEEE Symp. VLSI Technol.,* Jun. 2020, pp. 1-2, doi: 10.1109/VLSITechnology18217.2020.9265024

[19] M. H. Park, H. J. Kim, Y. J. Kim, T. Moon, and C. S. Hwang, "The effects of crystallographic orientation and strain of thin Hf$_{0.5}$Zr$_{0.5}$O$_2$ film on its ferroelectricity," *Appl. Phys. Lett.*, 10.1063/1.4866008.

[20] A. Kashir, H. W. Kim, S. Oh, and H. Hwang, "Large remnant polarization in a wake-up free Hf$_{0.5}$Zr$_{0.5}$O$_2$ ferroelectric film through bulk and interface engineering Orthorhombic HZO ~ 35 nm." *ACS Appl. Electron. Mater.* vol. 3, no. 2, pp. 629-638, Jan. 2021, doi: 10.1021/acsaelm.0c00671.

[21] S. Estandía, N. Dix, J. Gazquez, I. Fina, J. Lyu, M. F. Chisholm, J. Fontcuberta, and F. Sánchez, "Engineering Ferroelectric Hf$_{0.5}$Zr$_{0.5}$O$_2$ Thin Films by Epitaxial Stress," *ACS. Appl. Electron. Mater*, vol. 1, no. 8, pp. 1449–1457, Aug. 2019, doi: 10.1021/acsaelm.9b00256.

[22] Y. J. Lin, C. Y. Teng, C. Hu, C. J. Su, and Y. C. Tseng, "Impacts of surface nitridation on crystalline ferroelectric phase of Hf$_{1-x}$Zr$_x$O$_2$ and ferroelectric FET performance," *Appl. Phys. Lett.*, vol. 119, no. 19, Nov. 2021, doi: 10.1063/5.0062475.

[23] P. Yuan, B. Wang, Y. Yang, S. Lv, Y. Wang, Y. Xu, P. Jiang, Y. Chen, Z. Dang, Y. Ding, T. Gong, and Q. Luo, "Enhanced Remnant Polarization (30 µC/cm$^2$) and Retention of Ferroelectric Hf$_{0.5}$Zr$_{0.5}$O$_2$ by NH3 Plasma Treatment," *IEEE Electron Device Letters*, vol. 43, no. 7, pp. 1045–1048, Jul. 2022, doi: 10.1109/LED.2022.3178867.

[24] M. Klinger, "More features, more tools, more CrysTBox", *J. Appl.Cryst*. vol 50, no.4, pp. 1226-1234, 2017. doi: 10.1107/S1600576717006793.

[25] Y. H. Lee, S. D. Hyun, H. J. Kim, J. S. Kim, C. Yoo, T. Moon, K. D. Kim, H. W. Park, Y. B. Lee, B. S. Kim, J. Roh, M. H. Park, and C. S. Hwang, "Nucleation-Limited Ferroelectric Orthorhombic Phase Formation in Hf$_{0.5}$Zr$_{0.5}$O$_2$ Thin Films," *Adv. Electron. Mater*, vol. 5, no. 2, Feb. 2019, doi: 10.1002/aelm.201800436.

[26] H. A. Hsain, Y. Lee, G. Parsons, and J. L. Jones, " Compositional dependence of crystallization temperatures and phase evolution in hafnia-zirconia (Hf$_x$Zr$_{1-x}$)O$_2$ thin films ", *Appl. Phys. Lett*., vol. 116, no. 19, May 2020. doi: 10.1063/5.0002835

[27] V. Gaddam, D. Das, and S. Jeon, "Insertion of HfO$_2$ Seed/Dielectric Layer to the Ferroelectric HZO Films for Heightened Remanent Polarization in MFM Capacitors", IEEE Transactions on Electron Devices, vol. 67, no. 2, pp. 745-750, Feb. 2020, doi: 10.1109/TED.2019.2961208.

[28] A. Kashir, M. G. Farahani, and H. Hwang, "Towards an ideal high-κ HfO$_2$–ZrO$_2$-based dielectric", Nanoscale, 13(32), pp.13631-13640, June 2021. doi: 10.1039/D1NR02272E.

[29] U. Schroeder, M. Materano, T. Mittmann, P. D. Lomenzo, T. Mikolajick, and A. Toriumi, "Recent progress for obtaining the ferroelectric phase in hafnium oxide based films: Impact of oxygen and zirconium", Jpn. J. Appl. Phys., vol. 58, Nov. 2019. doi: 10.7567/1347-4065/ab45e3.

[30] R. Materlik, C. Kunneth, and A. Kersch, "The origin of ferroelectricity in Hf$_{1−x}$Zr$_x$O$_2$: A computational investigation and a surface energy model", J. Appl. Phys., vol. 117, Apr. 2015. doi: 10.1063/1.4916707.

[31] S. F. Lombardo, M. Tain, K. Chae, J. Hur, N. Tasneem, S. Yu, K. Cho, A. C. Kummel, J. Kacher, and A. I. Khan, "Local epitaxial-like templating effects and grain size distribution in atomic layer deposited Hf$_{0.5}$Zr$_{0.5}$O$_2$ thin film ferroelectric capacitors," *Appl. Phys. Lett.*, vol. 119, no. 9, Aug. 2021, doi: 10.1063/5.0057782.

[32] C.-H. Choi, L. Hultman, W.-A. Chiou, and S. A. Barnett "Growth of epitaxial TiN thin films on Si(100) by reactive magnetron sputtering," *J. Vac. Sci. Technol.*, vol. 9, no. 2, p. 221, Mar. 1991, doi: 10.1116/1.585597.

[33] J. Narayan, P. Tiwari, X. Chen, J. Singh, R. Chowdhury, and T. Zheleva, "Epitaxial growth of TiN films on (100) silicon substrates by laser physical vapor deposition," *Appl. Phys. Lett.*, vol. 61, no. 11, pp. 1290–1292, 1992, doi: 10.1063/1.107568.

[34] H. Wang, A. Tiwari, A. Kvit, X. Zhang, and J. Narayan, "Epitaxial growth of TaN thin films on Si(100) and Si(111) using a TiN buffer layer," *Appl. Phys. Lett.*, vol. 80, no. 13, pp. 2323–2325, Apr. 2002, doi: 10.1063/1.1466522.

[35] D. Reisinger, M. Schonecke, T. Brenninger, M. Opel, A. Erb, L. Alff, and R. Gross, "Epitaxy of Fe$_3$O$_4$ on Si(001) by pulsed laser deposition using a TiN/MgO buffer layer," *J. Appl. Phys*, vol. 94, no. 3, pp. 1857–1863, Aug. 2003, doi: 10.1063/1.1587885.

[36] R. Cao, Y. Wang, S. Zhao, Y Yang, X. Zhao, W. Wang, X. Zhang, H. Lv, Q. Liu, and M. Liu, "Effects of Capping Electrode on Ferroelectric Properties of Hf$_{0.5}$Zr$_{0.5}$O$_2$ Thin Films," *IEEE Electron Device Letters*, vol. 39, no. 8, pp. 1207–1210, Aug. 2018, doi: 10.1109/LED.2018.2846570.

[37] S. S. Fields, T. Cai, S. T. Jaszewski, A. Salanova, T. Mimura, H. H. Heinrich, M. D. Henry, K. P. Kelley, B. W. Sheldon, and J. F. Ihlefeld, "Origin of Ferroelectric Phase Stabilization via the Clamping Effect in Ferroelectric Hafnium Zirconium Oxide Thin Films", *Adv. Electron. Mater*., vol.8, no. 12, 2200601 Aug. 2022). doi: 10.1002/aelm.202200601.